\documentclass[11pt,a4paper]{article}

\usepackage{graphicx}
\usepackage{color}
\usepackage{tikz}
\usetikzlibrary{matrix}
\usepackage{amsmath,amssymb}

\usepackage{url}
\usepackage{breakurl}
\usepackage[breaklinks]{hyperref}

\newcommand{\R}{\mathbb{R}}

\usepackage{a4wide}
\usepackage{wrapfig}

\advance\topmargin by -13mm
\advance\oddsidemargin by -3mm
\advance\evensidemargin by -3mm
\advance\textheight by 23mm
\advance\textwidth by 3mm

\newcommand{\dwave}{D-Wave}
\newcommand{\CPLEX}{CPLEX}
\newcommand{\bb}{branch-and-bound}
\newcommand{\bc}{branch-and-cut}

\begin{document}

\title{Performance of a Quantum Annealer\\for Ising  Ground State Computations\\on Chimera Graphs}

\author{Michael J{\"u}nger\footnote{University of Cologne, Department of Computer Science, Germany, \texttt{mjuenger@informatik.uni-koeln.de}}
\and
Elisabeth Lobe\footnote{DLR German Aerospace Center, Braunschweig, Germany, \texttt{elisabeth.lobe@dlr.de}}
\and
Petra Mutzel\footnote{TU Dortmund University, Department of Computer Science, Germany, \texttt{petra.mutzel@cs.tu-dortmund.de}}
\and
Gerhard Reinelt\footnote{Heidelberg University, Department of Computer Science, Germany, \texttt{ip121@uni-heidelberg.de}}
\and
Franz Rendl\footnote{University of Klagenfurt, Department of Mathematics, Austria, \texttt{franz.rendl@aau.at}}
\and
Giovanni Rinaldi\footnote{Istituto di Analisi dei Sistemi ed Informatica, Rome, Italy, \texttt{giovanni.rinaldi@iasi.cnr.it}}
\and
Tobias Stollenwerk\footnote{DLR German Aerospace Center, Cologne, Germany, \texttt{tobias.stollenwerk@dlr.de}}
}

\maketitle             

\begin{abstract}
Quantum annealing  is getting increasing attention in combinatorial optimization. 
The quantum processing unit by \dwave\ is constructed to approximately solve 
Ising  models on so-called Chimera graphs. 
Ising  models are equivalent to quadratic unconstrained binary optimization (QUBO) problems and maximum cut problems on the associated graphs. 
We have tailored \bc\ as well as semidefinite programming algorithms  for solving Ising  models for Chimera graphs to provable optimality and use the strength of these approaches for  comparing our solution values to those obtained on the current quantum annealing  machine \dwave\ 2000Q.
This allows for the assessment of the quality of solutions produced by the \dwave\ hardware. It has been a matter of discussion in the literature how well the \dwave\ hardware performs at its native task, and our experiments shed some more light on this issue.

\end{abstract}

\section{Introduction}\label{sec:introduction}

The company 
\dwave\
has introduced a special hardware for realizing the quantum annealing  approach.
Their device basically finds approximate solutions of certain Ising  spin glass problems on Chimera graphs. 
Possible interactions between qubits
are defined by a Chimera graph whose nodes correspond to the qubits and whose edges
give the possible interactions. The machine, ``programmed'' by setting the values $J_{ij}$ for the edges of
the Chimera graph and $h_i$ for the qubits, finds
Ising  ($\pm1$) configurations with small energy by quantum annealing.

In order to solve a problem with the help of this machine, one must write a program on a conventional front end computer
that transforms a given problem instance into a sequence of Ising  problems on a Chimera graph. Since applications are rarely defined on
Chimera graphs, a minor-embedding has to be employed~\cite{cai2014}, which is a challenge by itself.
Moreover, depending on the mapping, the experimental results of the machine may be very different. 
So, here, an experiment on a classical and a quantum machine would compare two different things.
The \dwave\ machine gives a heuristic solution of the Ising  instances in a fraction of a second. 
It does not guarantee true optima, but conducting many runs on the same instance may increase the chances to find them~\cite{McGeochW13}. 

The outcome of many empirical comparisons was that the tested quantum annealers do not have enough qubits so far for solving 
large problems. 
In their very interesting report on benchmarking a smaller \dwave\ machine, Parekh  et al.~\cite{ParekhWS+16} predict that this will change if the number of qubits
reaches 2048, ``at which point exact classical algorithms will not longer be able to readily obtain optimal solutions for 
commonly used random QUBO benchmark instances on the Chimera graph.''
This was a motivation for us to exploit the ability of exact solvers for instances of larger sizes in order to assess the 
solution quality of the current \dwave\ 2000Q machine that has indeed 2048 qubits.

\paragraph{Our Contribution}

We transform the Ising  spin glass problem instances given to the \dwave\ hardware into maximum cut problem instances 
and then use an exact algorithm for solving them to optimality.
We use exact approaches based on \bc~\cite{BarahonaGJR88,DeSimoneDJ+95,DeSimoneDJ+1996,BonatoJRR14},  
semidefinite programming~\cite{RendlRW10},
and dynamic  programming
approaches~\cite{Selby13a,Selby14}. All of these approaches were tailored for Chimera graphs and
we compare the exact solution values to those obtained by the current \dwave\ 2000Q.

We compare on the lowest possible level in the sense that
we take our codes as simulators of the \dwave\ hardware in order to circumvent all the messy details that come in when  
solving a problem by transforming it to a sequence of Ising  instances for the \dwave\ hardware on the classical front-end. 
We consider this the only fair comparison, namely using universal approaches on both sides,
the quantum and the classical machine side.

This allows for the assessment of the quality of solutions produced by the \dwave\ hardware, rather than a compound of 
algorithms running on a conventional computer that may call an oracle for Ising  instances on Chimera graphs. 
It has been a matter of 
discussion in the literature how well the \dwave\ hardware performs at its native task, and our experiments shed some more 
light on this issue.

\section{Previous work}

Since QUBO problems can be (linearly) transformed into Ising  models,
the \dwave\ machine is capable of solving a wide class of optimization problems.
So far, quantum annealing  has been used for many combinatorial optimization problems such as, e.g.,
maximum independent set~\cite{ParekhWS+16,YarkoniPB18}, community detection~\cite{NegreUM19,ParekhWS+16}, 
clustering~\cite{BauckhageBCOSW18}, graph isomorphism~\cite{CaludeDH17}, and maximum 2-SAT~\cite{McGeochW13}.
Moreover, small real world application problems have been solved with the help of the \dwave\ quantum annealer 
\cite{rieffel2015,Venturelli2015,Perdomo-Ortiz2015,stollenwerkFGA2019,stollenwerkATM2019}.

It has been claimed that a \dwave\ Two quantum computer is $3600$ times faster than a conventional computer.
This statement goes back to a computational study of McGeoch and Wang~\cite{McGeochW13}. Although the paper does not
claim this in this generality, nevertheless,  a race has been started in which
researchers have compared the outcome of their classical approaches with those of using quantum annealing  by \dwave.
While McGeoch and Wang~\cite{McGeochW13} have used the \CPLEX\ quadratic programming software for solving the Ising  models, 
Dash and Puget~\cite{DashP15} have linearized the problem and solved the instances with the \CPLEX\ mixed integer linear 
programming software. With this better approach they could observe a speed-up factor of only~17.

Meanwhile, there exist many comparisons in which the quantum annealing  approach by
\dwave\ has been compared to classical algorithms.
Recently, many of them have been on combinatorial optimization problems.
Often, special purpose algorithms tailored to the considered optimization problem 
have been used in these comparisons with the result that the classical algorithms dominate the quantum 
annealing~\cite{BoixoRI+13}.

Parekh  et al.~\cite{ParekhWS+16} state that ``it is generally unknown which algorithm will perform best on a given
instance of a combinatorial optimization problem''~\cite{ParekhWS+16},
and therefore they suggest to compare on the mapped instances instead.
In this view, it does not seem to be fair to use a tailored algorithm and compare it with a universal algorithm.
A conclusion drawn that the quantum annealing  approach is worse would not be correct.
Parekh  et al.~\cite{ParekhWS+16} and Coffrin et al.~\cite{CoffrinNB17} have also suggested ideas to benchmark with quantum 
annealing.

Saket~\cite{Saket13} has suggested a PTAS for the Ising  problem on the Chimera graph. His algorithm approximates the value of 
minimum energy in a Chimera graph with $n$ nodes within a factor of $(1-\epsilon)$ in time $O(n2^{32/\epsilon})$. 
Selby~\cite{Selby13a,Selby14} 
provides an exact algorithm based on
dynamic programming that exploits the relatively low treewidth of Chimera graphs.

For our algorithms, we transform the Ising  problem to a maximum cut problem as described below.
The maximum cut problem has been shown to be NP-hard for general graphs by Karp~\cite{Karp72} and APX-hard by Papadimitriou
and Yannakakis~\cite{PapadimitriouY1991}.
Goemans and Williamson~\cite{GoemansW1995} have suggested a randomized constant factor approximation algorithm for instances with nonnegative weights that has 
been derandomized by Mahajan and Ramesh \cite{MahajanR1999} and has  performance guarantee 0.87856.
There are a number of special cases, for which the weighted maximum cut problem can be solved in polynomial time. Among them are 
planar input graphs~\cite{Hadlock75,ShihWuKuo90,LiersP12}, embedded 1-planar graphs~\cite{DahnKM18}, graphs not contractible 
to~$K_5$~\cite{Barahona83}, toroidal graphs~\cite{Barahona1983,GalluccioLoebl1998}, graphs without long odd 
cycles~\cite{GrotschelN84},  
and instances where the edges with positive weights make a graph with bounded vertex cover number~\cite{McRR03}.

Early approaches based on integer linear programming models have been suggested by Barahona et 
al.~\cite{BarahonaM86,BarahonaGJR88,Barahona1989}. 
A breakthrough for optimally solving Ising  spin glass models has been achieved in the 1990s by tailoring the algorithm to the 
special structure of grid graphs (see, e.g., \cite{DeSimoneDJ+95,DeSimoneDJ+1996}).
Based on this work there is a publically available spin glass solver~\cite{SpinglassSolver96}.
Whereas the cut polytope has been studied extensively for complete graphs~\cite{BarahonaM86}, only recently, new studies on 
lifting and separation procedures provided ideas for exploiting the structure of arbitrary graphs~\cite{BonatoJRR14}. Our 
work is heavily based on this previous work.

Rendl et al.~\cite{RendlRW10} have suggested an alternative approach based on semidefinite programming (SDP).
An SDP based \bb\ algorithm, the ``Biq Mac Solver -- Binary quadratic and Max cut Solver'' 
\cite{BiqMac09,RendlRW10} is publically available.
Approaches based on relaxations of the problem using eigenvalues have been suggested in~\cite{DelormeP93,PoljakR95}.

\section{Ground States of Ising  Spin Glasses}

A fundamental problem in statistical solid state physics is the determination of ground states of Ising  spin glasses.
We can think of an Ising  Model as a physical system in which the nodes, the \emph{spins},  of a graph $G=(V,E)$ represent particles (atoms) and the edges correspond to the interactions between the particles. The given interactions $J_{ij}$ describe the forces acting between two spins, and the node weights describe the forces acting on the nodes (e.g., the effect of some external magnetic field). A classical physical \emph{state} of the system describes the spin configurations for each spin which can be either up ($+1$) or down ($-1$).
Given a spin configuration $s=(s_1,s_2,\ldots,s_n)\in\{+1,-1\}^n$, 
\[H(s)=\sum_{ij\in E}J_{ij}s_is_j+\sum_{i\in V}h_is_i\]
gives the energy $H(s)$ of the configuration $s$.
The \emph{Ising  Spin Glass Ground State Problem (IM)} is to determine a \emph{ground state}, i.e., a configuration $s^*$ of minimum energy such that
\begin{equation}
H(s^*)=\min\bigl\{H(s)\mid s\in\{+1,-1\}^n\bigr\}.\label{eq:dwaveprob}
\end{equation}
If all node weights $h_i=0$ then we also talk about an \emph{IM with zero field}.

Beyond their importance in statistical physics, Ising  models have many applications, e.g., in chemistry (motion of atoms), neuroscience (activity of neurons), and the layout of electronic circuits \cite{BarahonaGJR88,DeSimoneDJ+95}.
The general IM has been shown to be NP-complete~\cite{Barahona1982},
Choi has also shown NP-completeness for Ising  models on Chimera graphs~\cite{Choi08}.

\section{\dwave\ Quantum Annealing}

The \dwave\ hardware is designed to find good approximate solutions to the Ising  Spin Glass Ground State Problem. In this machine
the spins correspond to the qubits. The possible interactions between qubits are represented by a so-called
\emph{Chimera graph}, i.e., a nonzero interaction $J_{ij}$ between qubits $i$ and $j$ is only possible if $ij$ is an edge in this graph.
Furthermore there is an external field that can act on each qubit $i$  (with an individual strength $h_i$).

\begin{wrapfigure}{R}{0.32\textwidth}
\begin{center}
\begin{tikzpicture}
\matrix (Chimera)[matrix of nodes,nodes in empty cells,nodes={inner sep=0mm,minimum width=4pt,outer sep=0pt,circle,thick,draw},column sep={6mm,between origins},row sep={3mm,between origins}]
{
& & & & & & &\\
& & & & & & &\\
& & & & & & &\\
& & & & & & &\\
|[draw=none]|&|[draw=none]|&|[draw=none]|&|[draw=none]|&|[draw=none]|&|[draw=none]|&|[draw=none]|&|[draw=none]|\\
& & & & & & &\\
& & & & & & &\\
& & & & & & &\\
& & & & & & &\\
|[draw=none]|&|[draw=none]|&|[draw=none]|&|[draw=none]|&|[draw=none]|&|[draw=none]|&|[draw=none]|&|[draw=none]|\\
& & & & & & &\\
& & & & & & &\\
& & & & & & &\\
& & & & & & &\\
|[draw=none]|&|[draw=none]|&|[draw=none]|&|[draw=none]|&|[draw=none]|&|[draw=none]|&|[draw=none]|&|[draw=none]|\\
& & & & & & &\\
& & & & & & &\\
& & & & & & &\\
& & & & & & &\\
};
\foreach \a in {1,...,4} { \foreach \b in {1,...,4}{ \foreach \c/\d in {1/2,3/4,5/6,7/8} \draw (Chimera-\a-\c) -- (Chimera-\b-\d); } }
\foreach \a in {6,...,9} { \foreach \b in {6,...,9}{ \foreach \c/\d in {1/2,3/4,5/6,7/8} \draw (Chimera-\a-\c) -- (Chimera-\b-\d); } }
\foreach \a in {11,...,14} { \foreach \b in {11,...,14}{ \foreach \c/\d in {1/2,3/4,5/6,7/8} \draw (Chimera-\a-\c) -- (Chimera-\b-\d); } }
\foreach \a in {16,...,19} { \foreach \b in {16,...,19}{ \foreach \c/\d in {1/2,3/4,5/6,7/8} \draw (Chimera-\a-\c) -- (Chimera-\b-\d); } }
\foreach \a in {1,...,4} { \foreach \c/\d in {2/4,4/6,6/8} \draw (Chimera-\a-\c) to [out=22,in=158]  (Chimera-\a-\d); }
\foreach \a in {6,...,9} { \foreach \c/\d in {2/4,4/6,6/8} \draw (Chimera-\a-\c) to [out=22,in=158]  (Chimera-\a-\d); }
\foreach \a in {11,...,14} { \foreach \c/\d in {2/4,4/6,6/8} \draw (Chimera-\a-\c) to [out=22,in=158]  (Chimera-\a-\d); }
\foreach \a in {16,...,19} { \foreach \c/\d in {2/4,4/6,6/8} \draw (Chimera-\a-\c) to [out=22,in=158]  (Chimera-\a-\d); }
\foreach \c in {1,3,5,7} {
  \foreach \a/\b in {1/6,2/7,3/8,4/9} \draw (Chimera-\a-\c) to [out=240,in=120] (Chimera-\b-\c); 
  \foreach \a/\b in {6/11,7/12,8/13,9/14} \draw (Chimera-\a-\c) to [out=240,in=120] (Chimera-\b-\c); 
  \foreach \a/\b in {11/16,12/17,13/18,14/19} \draw (Chimera-\a-\c) to [out=240,in=120] (Chimera-\b-\c); 
}
\end{tikzpicture}
\caption{Chimera Graph $C_4$}
\label{fig:C_4}
\end{center}
\end{wrapfigure}
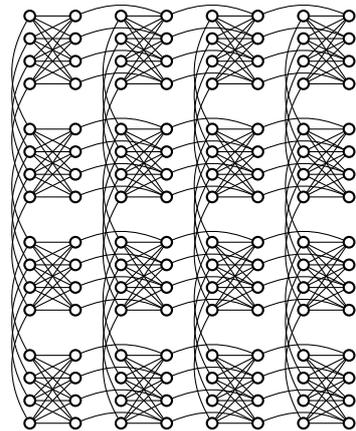

Rather than giving a formal definition of a Chimera graph, we look at $C_4$ shown in Figure~\ref{fig:C_4}. 
The subscript $k$ in $C_k$ specifies a $k\times k$ grid of $K_{4,4}$ subgraphs. The Chimera graph $C_k$ has $8k^2$ nodes and $24k^2-8k$ edges. If there is a non-zero field, it can be represented by an extra node and extra $8k^2$ edges associated with the $h_i$-values, so that we obtain a graph with $n=8k^2+1$ nodes and $m=32k^2-8k$ edges. 
The \dwave\ 2000Q  machine uses a $C_{16}$ Chimera graph having $2048$ nodes and $6016$ edges. Including the external field, we have 2049 nodes and 8064 edges.

By quantum annealing  the machine is capable of finding qubit (spin) configurations with low energy value.

As most real world problems do not conform to the Chimera structure, a minor embedding is needed to overcome this issue.
This may lead to two problems. 
First, for a fixed problem size the number of qubits is increased.
In the worst case of a complete graph problem, the \dwave\ 2000Q with its $C_{16}$ Chimera graph is limited to solving problems on $K_{64}$, the complete graph on 64 nodes.
For larger sizes, the problem can be partitioned on a classical computer and the parts are provided to the \dwave\ machine \cite{tran2016}.
Second, in order to implement the minor embedding, we need to force chains of qubits to be coupled together
by  assigning large negative artificial coupling weights on these. The absolute values of these weights must be large enough to prevent chain breaking. If the machine is not able to accommodate the required values, the solution might have broken chains in which case postprocessing methods like majority voting can be applied.
This might lead to a dramatic reduction of the success probability~\cite{stollenwerkATM2019}.

Another problem of the \dwave\ quantum annealer is its limited precision
for specifying the coefficients of the Ising  problem.
Thus, if two coefficients are too close to each other, they might be mapped to the same value and the problem is misspecified on the machine \cite{ParekhWS+16,stollenwerkATM2019}. 
More precisely, the \dwave\ hardware requires limited granularity of the coupling values $J_{ij}$ and field values $h_i$ in the following sense. Up to a scale factor, the values must be in the set \[\Gamma=\{-1,-1+\gamma^{-1},-1+2\gamma^{-1},\ldots,-\gamma^{-1},0,\gamma^{-1},2\gamma^{-1},\ldots,1-\gamma^{-1},1\}\] for some integer $\gamma\ge1$. In reality, $\gamma\approx30$. 
It is one of the tasks of the preprocessing routine to round coupling values outside $\Gamma$ to the nearest allowed value. When this happens, the machine does not solve the given instance but only an approximation of this instance.
This fact is often neglected in computational studies.

\section{Combinatorial Optimization using \dwave}

Any approach to a given combinatorial optimization problem using the \dwave\ hardware must run a program on the conventional front end that may call the \dwave\ hardware as an oracle for solving instances of type~(\ref{eq:dwaveprob}).

Two prominent combinatorial optimization problems are especially close to the Ising  problem~(\ref{eq:dwaveprob}): the maximum weight cut problem and the quadratic unconstrained binary optimization problem. The transformations are well known, see, e.g.~\cite{Barahona1989}. We define both problems and summarize the most important transformations for the purpose of this work.

\paragraph{Maximum Weight Cut Problem}

A cut in an undirected graph $G=(V,E)$ is defined by a subset $W\subseteq V$ and consists of the edges $\delta(W):=
\{ij\in E\mid i\in W,\,j\in V\setminus W\}$.
For edge weights $c_{ij}\in\R$ the \emph{Maximum Weight Cut Problem} or shortly \emph{MaxCut} is to determine a cut with maximum total weight:
\[\max\Bigl\{\sum_{ij\in\delta(W)}c_{ij}\mid W\subseteq V  \Bigr\}.\]

For the purpose of our exposition we prefer to represent edge sets $F\subseteq E$ by their characteristic vectors $\chi(F)\in\{0,1\}^E$ where $\chi_e=1$ if and only if $e\in F$. Then the problem reads
\begin{equation}
\max\Bigl\{\sum_{ij\in E}c_{ij}x_{ij}\mid x=\chi(\delta(W)) \mbox{ for }W\subseteq V\Bigr\}.\label{eq:maxcut}
\end{equation}

Solving~(\ref{eq:dwaveprob}) using a method for~(\ref{eq:maxcut}) is essential for our work and amounts to the following:
Let $G=(V,E)$ be the Chimera graph plus an extra special node $v$ with connecting edges to all qubit nodes. Let $c_{ij}=J_{ij}$ for all Chimera edges and $c_{iv}=h_i$ for all extra edges. Then an optimum $x^*$ for~(\ref{eq:maxcut}) with value $c^*$ gives rise to an optimum solution $s^*$ for~(\ref{eq:dwaveprob}) 
with value $\sum_{ij\in E}c_{ij}-2c^*$ in which $s^*_i=1-2x^*_{iv}$. 

Notice that the IM with zero field  is trivial for nonnegative interactions, because the Chimera graph is bipartite. Hence, the value of the maximum cut is equal to the sum of all edge weights (and therefore the ground state energy is the negative sum of all edge weights).

\paragraph{Quadratic Binary Optimization Problem}

The \emph{Quadratic Binary Optimization (QUBO) Problem} is
\begin{equation}
\max\Bigl\{x^TQx+q^Tx\mid x\in\{0,1\}^n\Bigr\}\label{eq:qubo}
\end{equation}
for a matrix $Q\in\R^{n\times n}$ and a vector $q\in\R^n$. Without loss of generality, we may assume that $Q$ is an upper triangular matrix with zero diagonal.

Using~(\ref{eq:dwaveprob}) for solving an instance of~(\ref{eq:qubo}) in which the nonzero entries $q_{ij}$ of the matrix~$Q$ correspond to couplers of the Chimera graph amounts to the following: 
Given $Q$ and $q$, set $J_{ij}=q_{ij}$ and $h_{i}=\sum_{j=1}^{i-1}q_{ji}+\sum_{j=i+1}^nq_{ij}+2q_i$. If $s^*$ is an optimum solution of~(\ref{eq:dwaveprob}) with value $H(s^*)$, then $x^*_{i} = (s^*_i+1)/2$ is an optimum solution for~(\ref{eq:qubo}) with value $(H(s^*)+\sum_{i=1}^{n-1}\sum_{j=i+1}^nq_{ij}+2\sum_{i=1}^nq_i)/4$. 

\paragraph{Granularity Issues}

With respect to granularity an instance of MaxCut with data values in $\Gamma$ is not affected, but even for  the closely related QUBO problem with all the $q_{ij}$- and $q_i$-values in $\Gamma$, the $h_i$-values resulting from the transformation above are not necessarily in $\Gamma$. 

In our experiments we restrict the values using  $\gamma=10$, such that we have \[\Gamma=\{-1.0,-0.9,\ldots,-0.1,0.0,0.1,\ldots,0.9,1.0\}\]
in order to be sure that we solve exactly the same instances in the quantum setting and our traditional mathematical optimization settings. 

\section{Solution of MaxCut to Optimality}

We review our approaches to solve MaxCut to optimality.

\subsection{Polyhedral Approach}

For $F\subseteq E$ we let $x(F):=\sum_{e\in F}x_e$. Then the integer linear program
\[\begin{array}{rcll}
\max c^Tx\\
x(Q)-x(C\setminus Q)&\le&|Q|-1&\mbox{for every cycle }C\mbox{ in }G\;\mbox{and all }Q\subseteq C,\,|Q|\mbox{ odd,}\\
x_e&\in&\{0,1\}&\mbox{for all }e\in E
\end{array}\]
models MaxCut, because the feasible solutions are exactly the characteristic vectors of cuts. The inequalities are called \emph{Odd Cycle Inequalities} and make sure that every cycle in $G$ intersects any cut in $G$ in an even number of edges.

If we replace ``$x_e \in \{0,1\}$'' by ``$0\le x_e\le1$'', we obtain a relaxation that gives an upper bound on the weight of any cut in $G$.
This relaxation can be solved by a cutting plane algorithm that starts without odd cycle inequalities and iteratively solves the linear program and subsequently determines odd cycle inequalities that are violated by the LP solution. 
We stop when no violated inequalities can be found. 
The problem of finding
violated odd cycle inequalities is solvable in polynomial
time~\cite{BarahonaM86} despite the fact that the number of cycles is exponential in general. Thus by~\cite{GroetschelLovaszSchrijver} the relaxation is
solvable in polynomial time. This procedure needs more than $|V|^3$ time, so for large instances, it is often advisable to run simpler (non-exact) procedures first and call the exact algorithm only if these fail. In the case
that $G$ is a Chimera graph, it is easy to enumerate all odd cycle inequalities on 3- and 4-cycles efficiently. 

If the solution of the relaxation is integral (which is always
the case when $G$ is planar), 
we have solved the MaxCut problem, otherwise we can embed the cutting plane algorithm in a \bb\ scheme. This approach is called \emph{\bc}.

The above relaxation can be strengthened by other inequalites that are valid for all characteristic vectors of cuts. For  details, see, e.g., \cite{LJRR2004} and~\cite{BonatoJRR14}. 
For this work, we have identified the strongest inequalities that are valid for all characteristic vectors of $K_{4,4}$-subgraphs with PANDA~\cite{panda}, and we use them in our computations. Finally,  the computations can be sped up by heuristic methods that produce odd cycle inequalities fast, yet do not guarantee to find any if there are violated ones. Also, general integer programming inequalities strengthen the relaxation, and they are applied by the state-of-the-art LP/IP-solvers that we use in our software.

\subsection{Semidefinite Programming Approach}

The semidefinite programming (SDP) approach for the MaxCut problem  exploits the close connection between (1) and (2).
Given an edge weighted graph $G$ on $n$ vertices, the weighted adjacency matrix $A=(a_{ij})$ of order $n$ is defined by setting
$a_{ij} = a_{ji} = c_e$ for edge $e=ij$ and $a_{ij} = 0$ otherwise.
We also need the \emph{Laplacian} corresponding to $A$, which is again  a symmetric matrix of order~$n$, defined by
\[L_{ii} = \sum_k a_{ik}, \quad L_{ij} = -a_{ij} \mbox{ for }i \not=j.\]
The cut $\delta(W)$, defined by $W \subseteq V$, can also be expressed through $s \in \{-1,1 \}^n$ by setting $s_i=1$ for $i \in W$ and 
$s_i=-1$ for $i \notin W$. Then $ij \in \delta(W)$ exactly if $s_i s_j=-1$. 
We use the following well-known semidefinite relaxation for MaxCut:
\[\max \Bigl\{ \frac{1}{4}\langle L,S \rangle\mid\operatorname{diag}(S)=e, S \succeq 0 \Bigr\}.\]
This is a linear program over the cone of positive 
semidefinite matrices and can be solved to fixed precision in polynomial time. 

The relaxation can be further tigthened by enforcing $S$ to satisfy certain \emph{hypermetric inequalities}. The simplest class of them are the 
 \emph{triangle inequalities}. Consider any $f \in \{1, 0, -1\}^n$  where exactly three entries are nonzero. It is clear that $|s^T f| \geq 1$ for any 
$s \in \{-1,1\}^n$, hence 
$(s^T f) (f^T s) \geq 1$, which translates into a linear inequality in $S$,
$\langle S , ff^T \rangle \geq 1$.
Allowing 5 or~7 entries in $f$ to be nonzero, we get the 5-clique and  7-clique hypermetric inequalities. We iteratively solve the semidefinite 
program, identify constraints which have become inactive,  and hence are dropped, and then look for new violated constraints of the 
types just described. Practical details how the resulting SDP are solved, and how violated constraints are iteratively added, can be found in 
\cite{RendlRW10}.

\subsection{Exact and Heuristic Approaches by Selby}\label{subsec:selby}

Selby~\cite{Selby13a,Selby14} provides an exact and a heuristic
algorithm for Chimera graphs based on dynamic programming. It is well
known that MaxCut can be solved to optimality by dynamic programming
in time $O(2^wn)$, where $w$ is the treewidth of the graph (see, e.g.,
\cite{BoJa00}). 
Therefore, for graphs with small treewidth, dynamic
programming can be an option as an exact optimization algorithm of
practical use. 
Based on the fact that the treewidth of a Chimera graph
$C_k$ is $4k$ (see, e.g.,~\cite{BoKR16}), and so it is relatively
small with respect to $|V|$, Selby  implemented an exact algorithm, whose
code is provided in~\cite{Selby13a}, that exploits a tree
decomposition of a Chimera graph. In~\cite{Selby14} he reports on such
an algorithm. The method is quite effective for small values of $k$,
since instances with $k\leq8$ could be solved regularly. However, for
bigger values of $k$ the algorithm becomes impractical for both
excessive time and memory requirements.  

We ran experiments with the algorithm on $C_8$ instances. Indeed, the
times are comparable with those of the other exact
algorithms. Sometimes they are the best among them.
However, in the current implementation~\cite{Selby13a}, only the ground state energy is computed, but not the ground state.

In~\cite{Selby14} also a heuristic algorithm is described that
exploits the ability of solving MaxCut on low tree-with graphs to
optimality but that can be used also for large Chimera graphs. The
algorithm is of a randomized type and is based on
the following
technique, called \emph{subgraph sampling}, inspired by earlier
work~\cite{FCBZ12,HaDe04}. 
Here we briefly outline the algorithm.

We are given an instance of a weighted graph $G=(V,E)$. We define an
ordered collection of subsets of $V$, i.e.,
${\mathcal{S}}=\{S_i\subset V\mid i=1,\ldots, t\}$ with the following
properties:
\begin{itemize}
  \item [a)] each $S_i$ is a large subset of $V$;
  \item [b)] the induced subgraph $G[S_i]$ is connected and has a
    treewidth bounded by a parameter $r$ that does not depend on
    $|V|$;
  \item [c)] $G$ is the union of the induced subgraphs $G[S_i]$, for
    $i=1,\ldots, t$.
\end{itemize}

\paragraph{The inner algorithm}

Given a weighted graph $G=(V,E)$, a node subset $S$ with treewidth at
most $r$, and a spin configuration of the nodes in $V\setminus S$, the
algorithm finds the spin configuration of the nodes in $S$ that yields
the maximum cut among all spin configurations of the nodes in $S$.

The configuration of the nodes in $V\setminus S$ is never changed by
the algorithm. These nodes act as an external magnetic field on the
nodes in $S$. If parameter $r$ is small, the algorithm if pretty fast
in finding the optimal solution even if $S$ has a large size. The
algorithm is essentially deterministic; however, to break ties, a
random permutation is generated at the beginning that is used to
ensure the uniqueness of the optimum.

\paragraph{The outer algorithm}

Given a weighted graph $G=(V,E)$, a node subset $S$ with treewidth at
most $r$, and a spin configuration of the nodes in $V$, the algorithm
generates a random ordered collection of $t$ subsets of $V$,
$\{S_i\subset V\mid i=1,\ldots, t\}$. Then, for every
$i\in1,\ldots,t$, it executes the inner algorithm for $S_i$ using the
current spin configuration to set the one of the nodes in $V\setminus
S_i$. The solution generated by the inner algorithm becomes the
current spin configuration used for the next run.

\paragraph{The heuristic}

The heuristic starts by generating a random spin configuration of the
nodes in $V$. Then the outer algorithm is executed repeatedly. Between
two consecutive runs either a new spin configuration is randomly
generated afresh or the best configuration obtained in the previous
run is partially randomly perturbed. The latter is the default
option. In the default setting the random perturbation is obtained by
randomly generating a configuration for the nodes of 20\% of the
$K_{4,4}$ Chimera subgraphs, randomly chosen.

The algorithm is run until a certain condition is satisfied. For
example, in our experiments, the algorithm was stopped when the elapsed
wall clock time exceeded a prescribed duration.

\paragraph{How the induced subgraphs are chosen}

A Chimera graph $C_k$ (see Figure~\ref{fig:C_4}) can be thought of as
a square grid of $k\times k$ cells, each of them being a copy of a
$K_{4,4}$ graph. The 4 nodes on the right side of each cell are
connected to the corresponding nodes of the cell at its right in the
grid (except for the rightmost cell, of course). We call these nodes
the \emph{h-nodes}. The 4 nodes on the left side of each cell (except
those in the bottom row of the grid) are connected to the
corresponding nodes of the cell immediately below. These nodes are
called the \emph{v-nodes}. We can now define the node set
$H(w,0,i,j)\subset V$ as follows: we take the full node set $V$ and we
remove all h-nodes of the grid columns with index $c\equiv i
\pmod{w+1}$, except those that belong to grid row $j$ (row and column
indices in the grid start from~0). By interchanging the roles of rows
and columns and by replacing h-nodes with v-nodes, we define the node
subsets $H(w,1,i,j)\subset V$. There are~$2w$ such subsets; the
subgraph induced by each of them is connected and has treewidth equal
to $4w$.

In the Selby  heuristic the random ordered collection is built by
generating three random numbers: $x$ in the set $\{0,1\}$, $y$ in the
set $\{0,w+1\}$, and $z$ in the set $\{0,k-1\}$. Then the collection
is given by:
\[
H(w,x,y,z), \ldots, H(w,x,y+w,z), H(w,1-x,y,z), \ldots,
H(w,1-x,y+w,z),
\]
where the third argument is taken $(\bmod{w+1})$.

\section{Experiments}

Our MaxCut \bc\ code comes in two versions: For dense instances, we use the version \texttt{chimeramaxcut} that is tailored to a Chimera graph ($C_8$ or $C_{16}$) with or without field. For the sparse instances, we apply a reduction to a general smaller maxcut instance by removing isolated qubit nodes (no nonzero coupler/edge) as well as zero weight coupler- and field-edges.  We solve the latter using a MaxCut \bc\ code \texttt{generalmaxcut} for general graphs. The advantage is that the instance is smaller, the drawback is that we cannot use the special structural knowledge of the Chimera graphs. In \texttt{chimeramaxcut}, instead we fix all edge variables connecting isolated qubit nodes to zero.
We solve the linear programming relaxations with \CPLEX~\cite{cplex}
in \texttt{chimeramaxcut} and Gurobi~\cite{gurobi} in \texttt{generalmaxcut}.
Our SDP-solver is \texttt{sdpmaxcut}.
It uses MATLAB with interfaces in C. For all experiments,  \texttt{chimeramaxcut}
ran under the Debian~7 operating system on two 3.00GHz Intel Xeon CPUs E5-2690v2
with 10 cores each, i.e., on a total of 20 cores,
\texttt{generalmaxcut} ran on a MacBook Pro with a 2.9 GHz Core i7 processor,
and \texttt{sdpmaxcut} ran on a 3.1GHz Intel Optiplex 790 with  4 cores.

The running times of our three MaxCut codes do not matter for the purpose of this study. We invested up to 2 days of computation time in order to solve an instance to optimality. When this failed, we stopped the computation and returned the best spin configuration $\hat{s}$ along with its energy $H(\hat{s})$ and the best lower bound for the ground state energy determined at this point in time.

In our implementation of the Selby  heuristic we used all the default
settings of the code given in~\cite{Selby13a}. In particular,
for the parameter $w$ of the previous paragraph we chose value~3,
which produces subgraphs of treewidth 12. Consequently, for instances
of $C_{16}$, each inner algorithm solves a MaxCut problem to
optimality on an induced subgraph having
more than 88\% of the nodes in $V$.
We slightly modified the original implementation by making the code
suitable to be run in a multithread environment. Moreover, to make the
experiments reproducible in any computational setting, we replaced the
standard random number generator with the one proposed by
Knuth~\cite{knuth93}. Starting from the same seed, this generator
produces the same sequence no matter in which hardware it is run or
with which compiler it is built. A run of the heuristic for a single
instance consists of 80 parallel threads each initialized with a
different seed in the set $\{4711,\ldots,4790\}$. The wall clock time
allowed to the parallel algorithm is 30 seconds for all instances.

For the \dwave\ quantum annealer we will always consider the best solution of multiple runs.
For our experiments, we used $10000$ annealing  runs, no gauges and an annealing  time of $T_{\text{anneal}} = 20 \mu s$.
The time-to-solution with $99 \%$ certainty can be estimated by
\begin{equation*}
	T_{99} = \frac{\ln(1 - 0.99)}{\ln(1-p)} T_{\text{anneal}}
\end{equation*}
where $p$ is the success probability. It is estimated by the number of optimum solutions found divided by the total number of annealing  runs. 
One can relax this measure by allowing for suboptimal solutions up to a
certain deviation from optimum solutions.
In this case one would count the number of solutions which deviate up to
a certain percentage from the optimum solutions.
With this, we could estimate the success probability for finding a
solution within these limits.

All instances respect a fixed list of 17 faulty qubits and 2 faulty Chimera couplers of the \dwave\ 2000Q machine installed at the NASA Quantum Artificial Intelligence Laboratory at NASA Ames on which the ``\dwave'' experiments were carried out.

The instances used in this study are available at \url{https://informatik.uni-koeln.de/public/mjuenger/chimera-article-and-data}.

\subsection{McGeoch-Wang Instances}

Our first test  is on $C_8$-instances (\dwave\ Two) described in McGeoch and Wang~\cite{McGeochW13} and used by Dash and Puget~\cite{DashP15}. 
Here, the $\pm1$ edge weights (for Chimera edges and field edges) 
are drawn uniformly at random.

While the \dwave\ Two has 512 qubits, the instances have only 439 qubits. The original instances are not available to us. We generated 100 random instances in which 73 qubits are declared ``faulty''. In the northwest quarter of the \dwave\ 2000Q that we use in our experiments,  5 qubits are indeed faulty, and the other 68 qubits are chosen at random, as well as the $\pm1$ weights on all edges ($+1$ and $-1$ with equal probability) that are not connected to real or fake faulty qubits. So other than in the original experiment, not only the weights but also the faulty qubits are chosen at random. This seems reasonable since we do not know the fixed set of 73 ``faulty'' qubits in the original experiments.

Of course, for every instance, the given data can be interpreted as a QUBO instance rather than an Ising  instance. (The $J_{ij}$ form the matrix $Q$ and the $h_i$ form the vector $q$). However, doing so makes a hard problem very easy, as can be observed when studying the computational experiments in~\cite{DashP15}. We can confirm this and refrain from according experiments.

Indeed, as McGeoch and Wang, who performed only 1000 annealing  runs on \dwave\ Two, claim, our results on \dwave\ 2000Q are almost always optimum: 
Only in 5 of the 100 instances, the energy values differ from the optimum. 
The maximum relative deviation from the true ground state energy is $0.254$\%.

\begin{figure}[ht]
\begin{center}
\includegraphics[scale=.3]{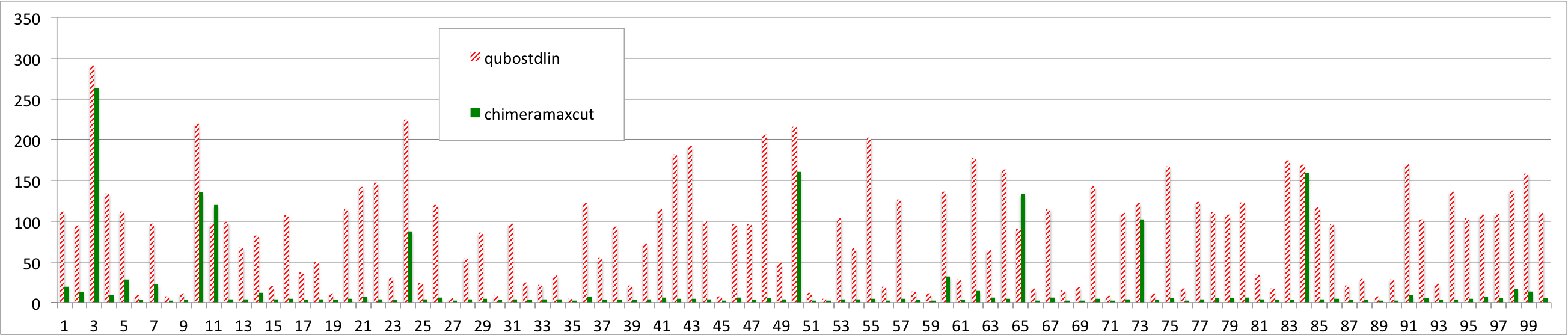} 
\caption{Running Times for the McGeoch-Wang instances in cpu seconds}
\label{fig:table2-cpu}
\end{center}
\end{figure}

The running times (in cpu seconds) of our re-implementation  ``\texttt{qubostdlin}'' of the standard linearization method used in~\cite{DashP15} and with  \texttt{chimeramaxcut} 
are given in Figure~\ref{fig:table2-cpu}. They are 
in the range $[4.42,291.08]$ with average $86.98$ for \texttt{qubostdlin} and in the range $[2.12,263.10]$ with average $16.75$ for \texttt{chimeramaxcut}.
For 92 of the 100 instances branching was not needed in \texttt{chimeramaxcut}.

We extended the experiments with additional runs on $C_{8}$ that respect only the 5 really faulty qubits in the northwest quarter of the \dwave\ 2000Q, giving instances on 507 qubits. The running times 
for  \texttt{qubostdlin} 
are between 307s and $15.7$h, $4.6$h on average, and for \texttt{chimeramaxcut} they are between 6s and 512s, 195s on average.
The cpu times on these harder instances clearly indicate that our specially tailored \bc\ imple\-men\-tat\-ion  \texttt{chimeramaxcut} outperforms the standard linearization method \texttt{qubostdlin} as has been predicted in~\cite{DashP15}.
\dwave\ finds true ground states for all 20 instances.

\begin{figure}[ht]
\begin{center}
\includegraphics[scale=.3]{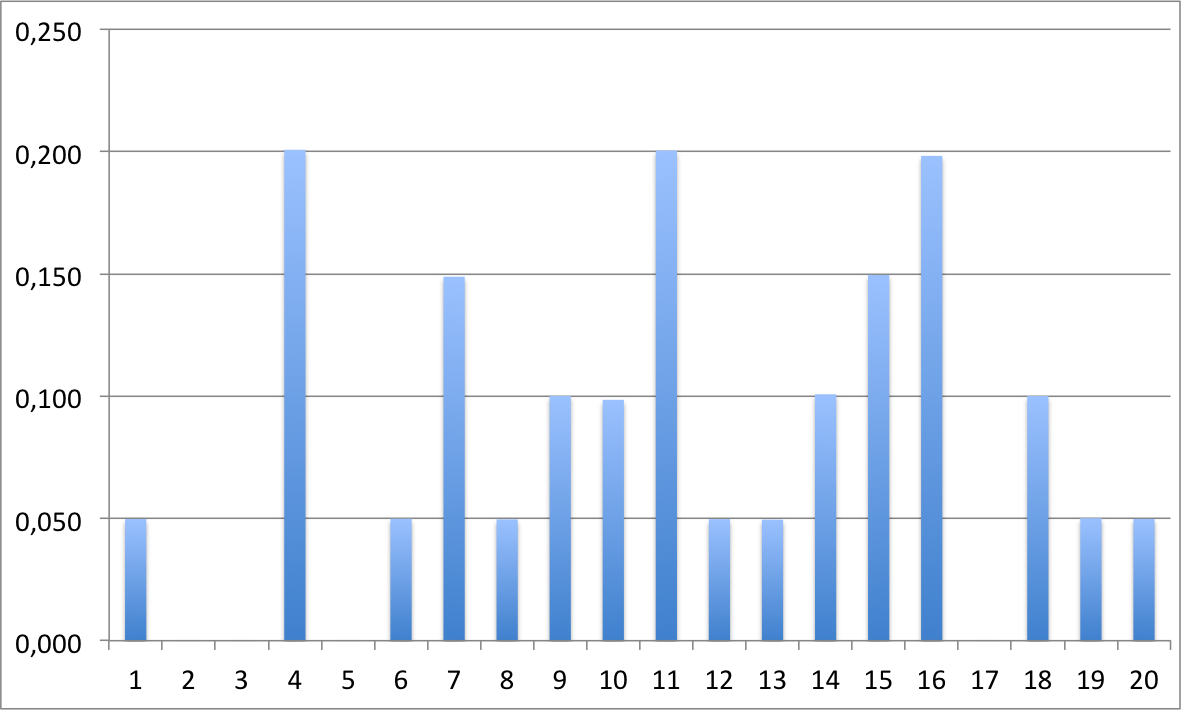}  
\caption{Relative deviations of the best \dwave\ energies from the best \texttt{chimeramaxcut} energies for the McGeoch-Wang-like instances on $C_{16}$ in percent}
\label{fig:c16mgw-gap}
\end{center}
\end{figure}

Finally, we extended the experiments with additional runs on large instances for $C_{16}$ only in direct comparison between \dwave\ and \texttt{chimeramaxcut}. Here, only the 17 faulty qubits and the 2 faulty couplers are respected, so that the instances are on 2031 qubits and 7950 nonzero coupler and field values.
The program \texttt{chimeramaxcut}  failed to compute proven optimum solutions in all but two cases  in reasonable computation time. Therefore, we limited the computation time to 10h for each instance and report only the energies of the best solutions found in that time frame along with lower bounds for the ground state energies.
In 16 out of 20 runs, the  \texttt{chimeramaxcut}  solutions are better than the \dwave\ solutions. 
Figure~\ref{fig:c16mgw-gap} shows the relative deviations of the best \dwave\ energies from the best \texttt{chimeramaxcut} energies in percent.

The relative deviation of the best \dwave\ energy from the best \texttt{chimeramaxcut} energy is at most $0.201$\% and $0.082$\% on average.

In summary, we could verify the statement in \cite{McGeochW13} for the 439-qubit-$C_8$-instances  that the quantum annealer almost always solves such instances to optimality (95\%), but this is not true for bigger instances in the sense that in 
80\% of the runs the \dwave\ energy for the $C_{16}$-instances is definitely not the ground state energy. However, the gaps are very small: at most $0.201$\%.

\subsection{Random Instances with Full Range within Granularity}

\begin{figure}[ht]
\begin{center}
\includegraphics[scale=.3]{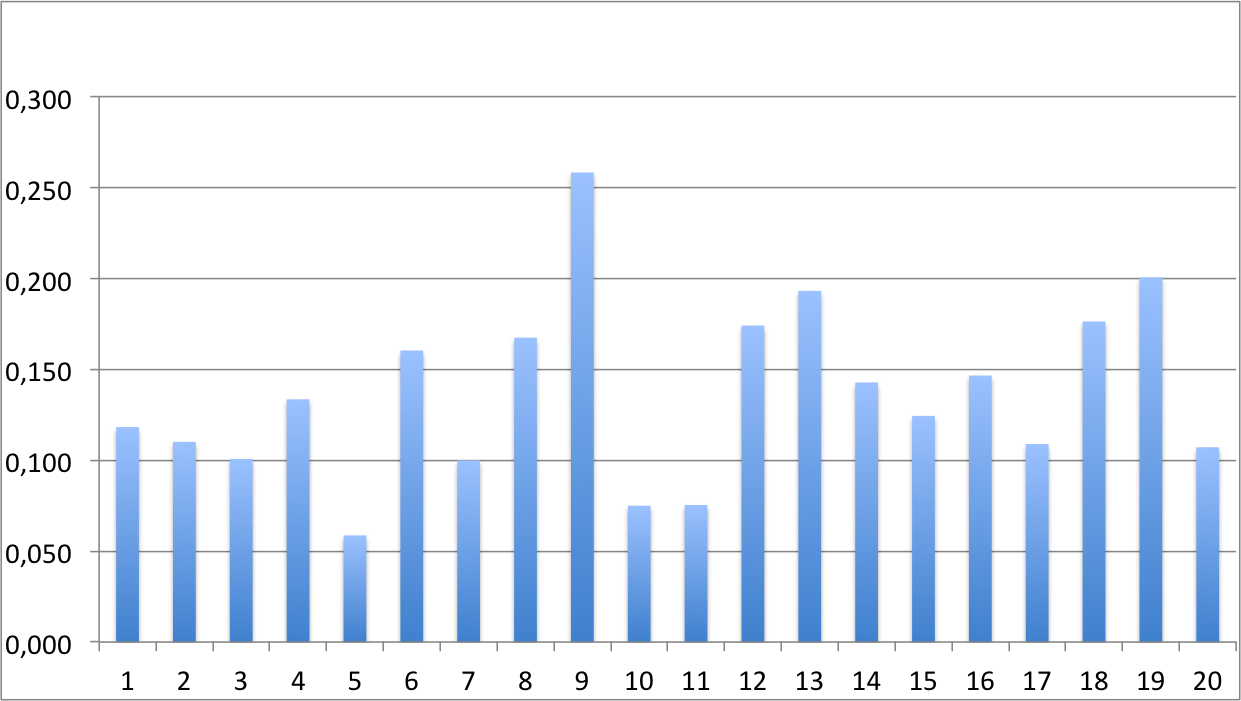}
\caption{Relative deviations of the best \dwave\ energies from true ground state energies for the random instances with full range within granularity on $C_{16}$ in percent}
\label{fig:afi-gap}
\end{center}
\end{figure}

The next experiment gives uniform random weights to all Chimera- and all field-edges within the entire spectrum 
$\Gamma=\{-1.0,-0.9,\ldots,-0.1,0.0,0.1,\ldots,0.9,1.0\}$. 
The relative deviations of the \dwave\ energies from the proven ground state energies are given in Figure~\ref{fig:afi-gap}. The maximum is $0.258$\% and the average is $0.137$\%.
The  \texttt{chimeramaxcut}  computation times are between 2m and 6m, about $2.5$m on average.
\dwave\ fails to compute a ground state for all 20 instances, with relative deviations between $0.059$\% and $0.258$\%.
Again, all \dwave\ energies are very close to the ground state energies.

\subsection{Selby  Instances}

Dash and Puget~\cite{DashP15} have reported that McGeoch-Wang-like instances with zero field, i.e., $h_i=0$ for all qubits, are very hard for ILP-based approaches. Our experiments have confirmed this. 
Selby~\cite{Selby17} states that 
instances in which the field is zero and the weights of the edges connecting different  $K_{4,4}$s are about twice as large as those within the $K_{4,4}$ subgraphs are
even harder for his approaches and also for \CPLEX.
We generated such instances on $C_8$ and $C_{16}$ in order to figure out the limits of our current exact approaches.

We chose uniform random weights in $\{-0.5,\ldots,-0.1,0.0,0.1,\ldots,0.5\}$ for the intra-$K_{4,4}$-edges and  in  $\Gamma=\{-1.0,\ldots,-0.1,0.0,0.1,\ldots,1.0\}$ for the inter-$K_{4,4}$-edges. The field is 0.

\begin{figure}[!ht]
\begin{center}
\includegraphics[scale=.3]{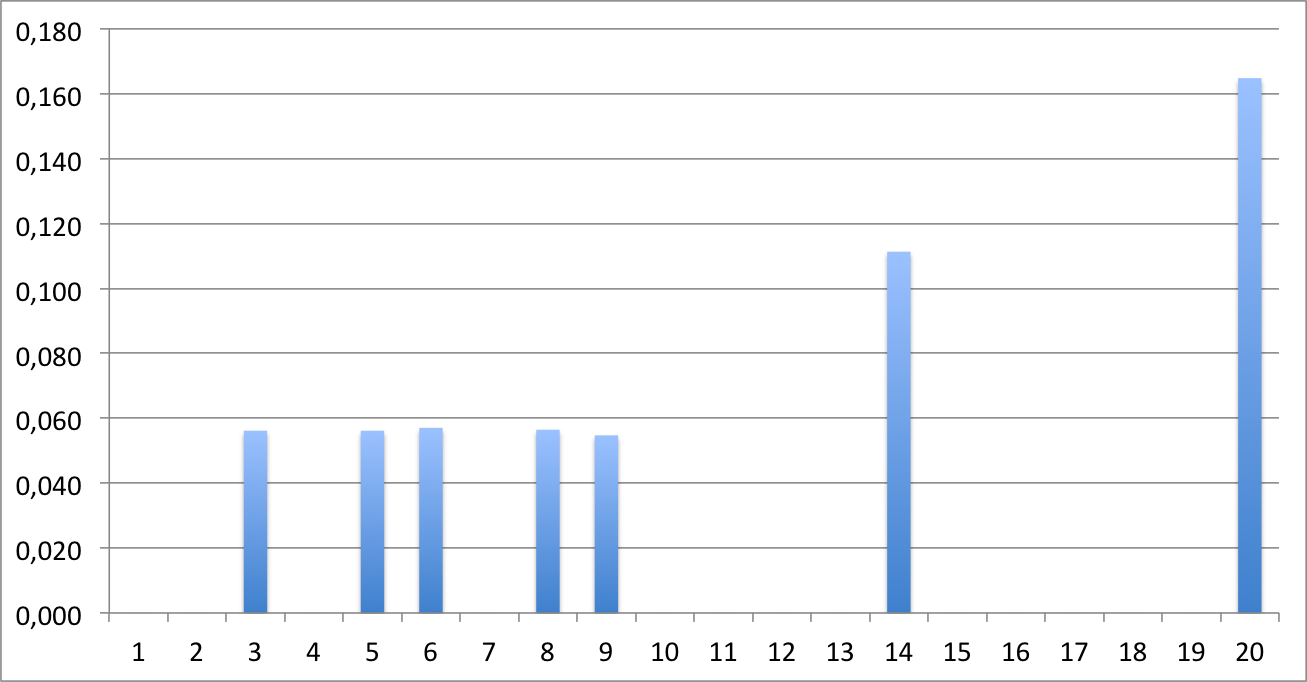}  
\caption{Relative deviations of the best \dwave\ energies from true ground state energies for the Selby  instances on $C_{8}$ in percent}
\label{fig:c8selby-gap}
\end{center}
\end{figure}

For the $C_8$-instances  \texttt{chimeramaxcut}  is not able to find a proven ground state within 10h in 6 of 20 cases, it takes between $0.5$h and $2.5$h for the others. The sdp code \texttt{sdpmaxcut} solves all but 1 instance to proven optimality within 5m and 15m. 
For the 6th instance, \texttt{sdpmaxcut} returns a spin configuration with energy $-350.6$ along with a lower bound for the ground state energy of $-351.2$. Using the exact Selby  optimizer, we could verify that $-350.6$ is indeed the ground state energy.
\dwave\ finds a true ground state in 13 of the 20 cases, the relative deviation  from the best \texttt{sdpmaxcut} energies is always less than $0.165$\%, see Figure~\ref{fig:c8selby-gap}. 

\begin{figure}[ht]
\begin{center}
\includegraphics[scale=.3]{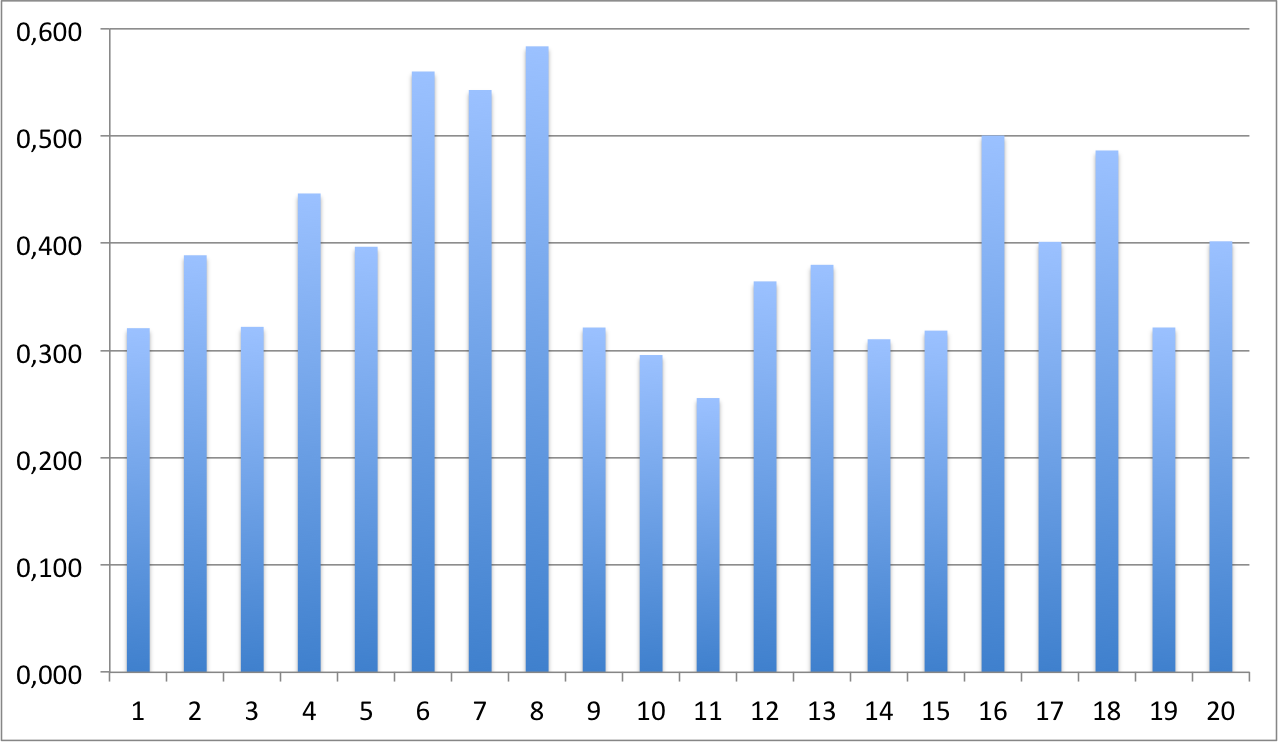}   
\caption{Relative deviations of the best \dwave\ energies from the best known energies for the Selby  instances on $C_{16}$ in percent}
\label{fig:c16selby-gap}
\end{center}
\end{figure}

The results for the $C_{16}$-instances are shown in Figure~\ref{fig:c16selby-gap}. 
None of our exact approaches is able  to compute proven optimum solutions for the $C_{16}$-instances in reasonable computation time. Therefore, we limited the computation time to 10h for each instance and return only the energies of the best solutions found in that time frame along with lower bounds for the ground state energies. We observe that the gaps are much higher (up to $0.584$\%) than those for the $C_8$-instances. 

\subsection{Maximum Independent Set Instances}

In order to identify hard instances, we also tried the Maximum Independent Set Instances (MIS) instances. 
Parekh  et al.~\cite{ParekhWS+16} have reported on a variety of computational experiments in which they compared exact as well as heuristic approaches with the quantum annealing  approach of \dwave\ Two (512 quantum bits). They found that their instances arising from the maximum independent set problem have been very hard to solve for the tested approaches.

They  suggested to set the weights of all coupler edges in the corresponding Ising  graph to $J_{ij}=1$ with probability $p=1/2$ and $J_{ij}=0$ otherwise. In order to get maximum independent set instances, the linear term values need to be the sum of the weights of the adjacent edges minus 2. Because of the limited precision we scale the values down so that our edges get weight $1/10$ and the linear term gets the value $\sum_{j:ij\in E}J_{ij}-0.2$. 

\begin{figure}[ht]
\begin{center}
\includegraphics[scale=.3]{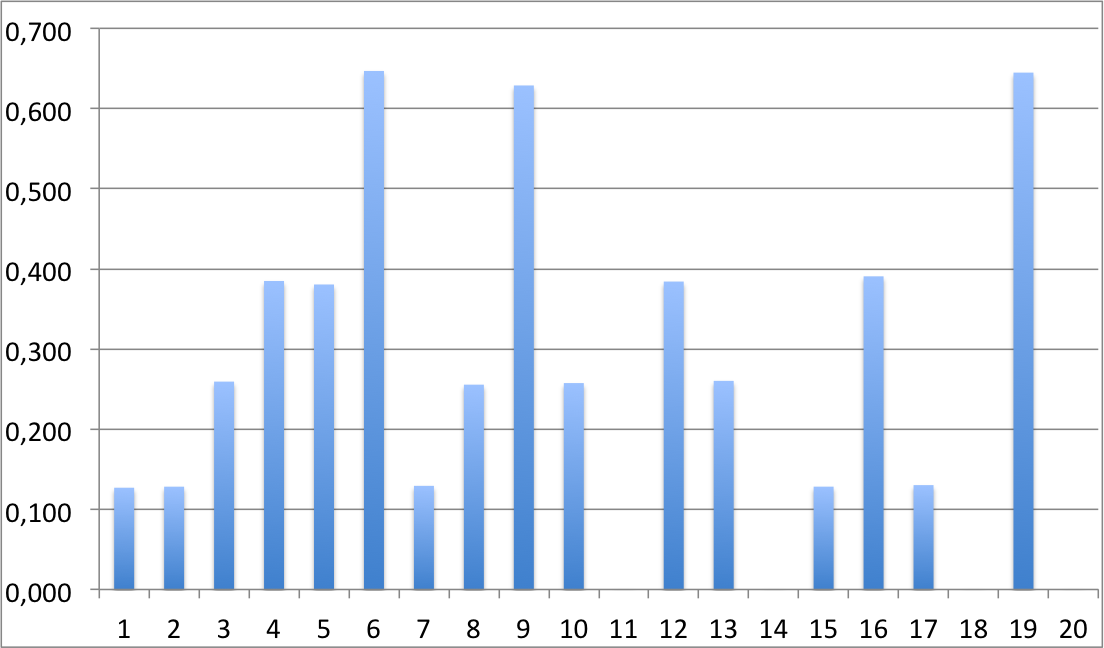} 
\caption{Relative deviations of the best \dwave\ energies from true ground state energies for the maximum independent set instances on $C_{16}$ in percent}
\label{fig:mis-gap}
\end{center}
\end{figure}

The results on solution quality are shown in Figure~\ref{fig:mis-gap}. 
Due to the sparsity of the instances, \texttt{generalmaxcut}  outperforms all other exact solvers by far. Solution times to proven optimality are between 1s and 3s.
\dwave\ identifies a correct ground state in 4 of 20 cases, for the others the relative energy deviation ranges from $0.127$\% to $0.647$\% from the true ground states.

This is the only case we encountered in which an instance set is easy for our exact traditional methods and hard for \dwave\ in the sense that true ground states are not found for the majority of the instances.

\subsection{Practical Instances from Gate Assignment in~\cite{stollenwerkFGA2019}}

These instances stem from a study of solving a real world problem from air traffic management with a \dwave\ quantum annealer~\cite{stollenwerkFGA2019}.
The problem is to find an optimum assignment of flights to gates at an airport such that the transit time for all passengers is minimum.
The $1035$ problem instances are exactly those used in the study after embedding them to the \dwave\ quantum annealer with \dwave's embedding heuristic \cite{cai2014}.
Before embedding, the original problem instances had between $6$ and $21$ binary variables. 
Different from all other instance sets in this study, these instances have weights that do not respect the $\gamma=10$ granularity.

\dwave\ finds true ground states for all instances, as confirmed by \texttt{generalmaxcut}  within at most $1.27$s and on average in about $0.03$s.

\subsection{Large Gate-Assignment-like Instances}

The instances used in \cite{stollenwerkFGA2019} are quite small. 
Their size is bounded from above due to increasing granularity with increasing problem size.
In order to generate instances which are close to the real world problem, but larger in size, we did the following.
We used the same method and data for extracting flight gate assignment problem instances from real world data as it was described in \cite{stollenwerkFGA2019},
i.e., use the connected components of the transfer passenger graph and apply binning on the passenger and time values.
In contrast to \cite{stollenwerkFGA2019} however, we restricted ourselves to larger instances with $50$ to $90$ binary variables.
After embedding with \dwave's heuristics \cite{cai2014}, these instances exhibit granularities which exceed the precision of the \dwave\ machine.
Therefore we employed binning after embedding once again to achieve $\gamma=10$ granularity.

For all 30 instances,  \texttt{chimeramaxcut} failed to compute ground states, but \texttt{sdpmaxcut} succeeded for 24 instances. For the remaining 6 instances,  \texttt{sdpmaxcut} found spin configurations whose energies deviate from the ground state energy by at most $0.09$\%. Computation times were between 3.6m and 4.3h.  

\begin{figure}[ht]
\begin{center}
\includegraphics[scale=.3]{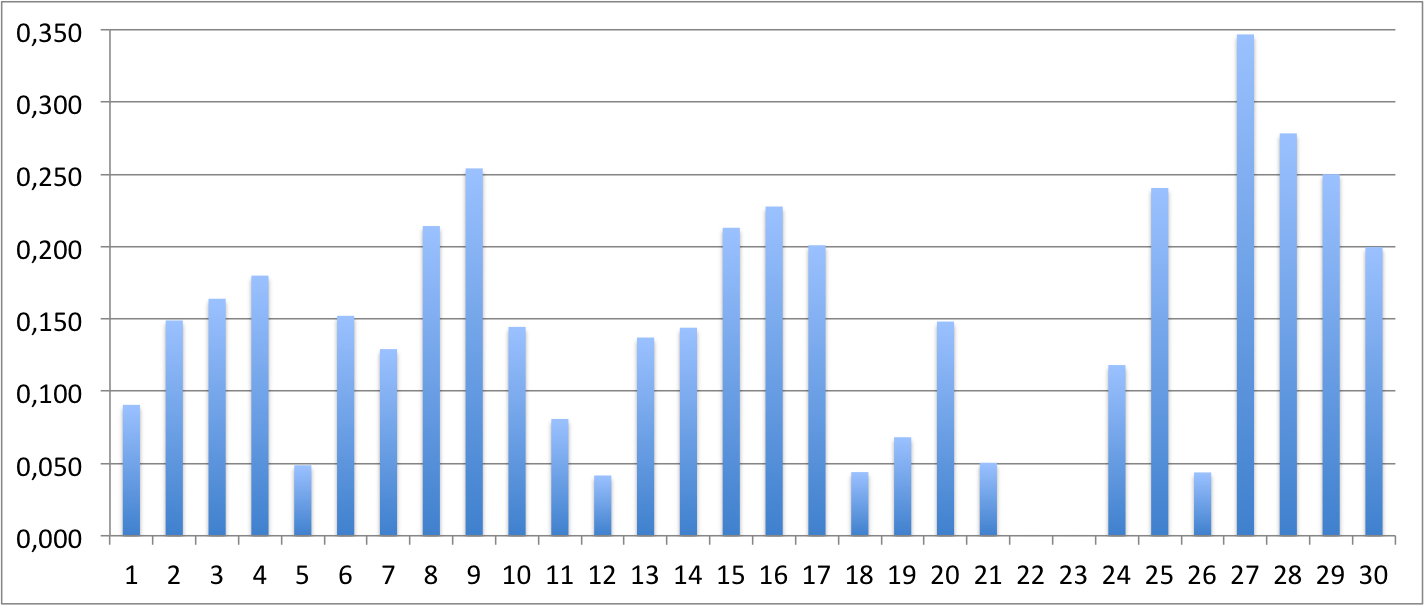} 
\caption{Relative deviations of the best \dwave\ energies from the best known energies for the large gate-assignment-like instances on $C_{16}$ in percent}
\label{fig:gal-gap}
\end{center}
\end{figure}

\dwave\ finds true ground states for 2 of the 30 instances, the relative deviation from the best known energy is at most $0.347$\% and $0.145$\% on average%
, see Figure~\ref{fig:gal-gap}. 

\subsection{Embedded $K_{64}$ Ising  Instances}

The interaction graphs of practical instances usually do not correspond to the restricted Chimera structure. 
Therefore, in nearly all cases it is required to embed the original problem into the hardware graph by mapping each original node to several qubits.
Since embedding an arbitrary graph into the Chimera graph is most likely a hard problem itself, a precomputed generic embedding for a complete graph can be used. 
The current \dwave\ architecture allows at most 64 fully connected nodes.
The resulting embedded instances have a specific structure including long and strongly coupled qubit chains.

To simulate these kinds of instances we have randomly generated Ising  instances on 
 graphs with 64 nodes. All edges in the random graph received a weight of either $+1$ or $-1$.
Since the value of a node $i$ whose weight exceeds the influence of the incoming edges, that is if
\[|h_i| \geq \sum_{j \in N(i)} |J_{ij}|,\] 
can be directly set to $-\operatorname{sign}(h_i)$, we avoided those preprocessable nodes by choosing random node weights from $\{-|N(i)| + 1,\ldots, |N(i)| - 1\}$ for each node $i \in V$, where  $N(i)$ denotes the set of neighbors of $i$.
Afterwards the resulting Ising  model has been transferred into the Chimera graph via two different embeddings.
If the construction of one of the embedded Ising  models was not possible using just integer coefficients between $-10$ and $10$ the corresponding original instance was rejected.
This was nearly always the case when the density of the random graph exceeded $0.25$.
Therefore, we have chosen the edge probabilities $0.23,0.24,0.25$ for generating the random instances.
In this way we created 40 $K_{64}$ Ising  instances, therefore 80 embedded ones. 
With edge weights divided by~10 they result in an instance set respecting our granularity restriction. 

\begin{figure}[ht]
\begin{center}
\includegraphics[scale=.3]{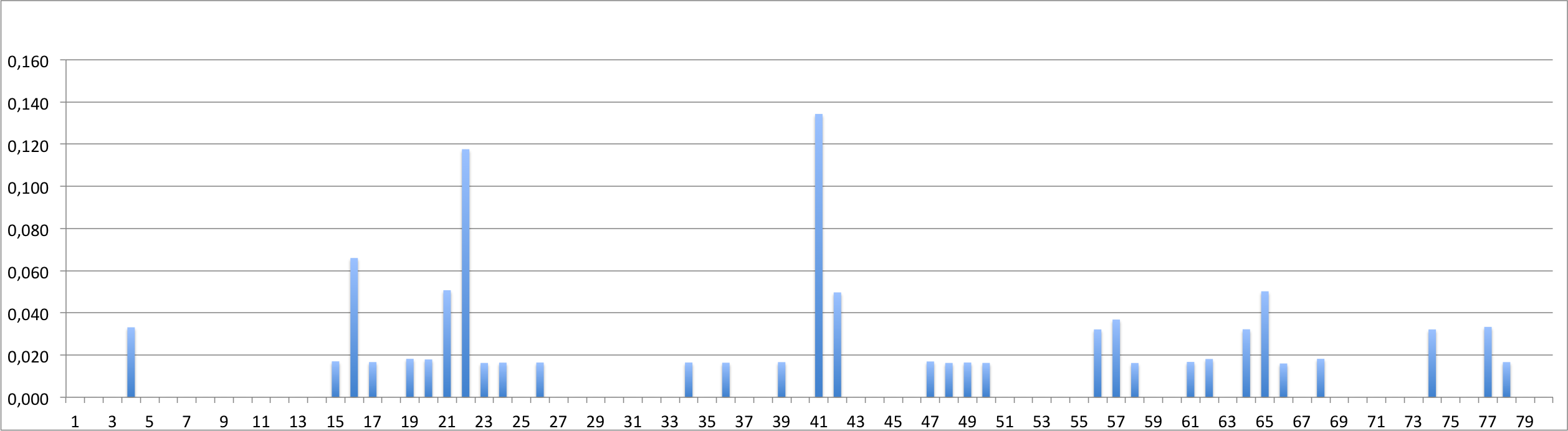} 
\caption{Relative deviations of the best \dwave\ energies from true ground state energies for the embedded $K_{64}$ Ising  instances on $C_{16}$ in percent}
\label{fig:k64ising-gap}
\end{center}
\end{figure}

\dwave\ finds true ground states for 48 of the 80 instances, as confirmed by  \texttt{generalmaxcut}  within at most $2.18$s and on average $0.92$s, the relative deviation from the ground state energy is at most $0.134$\%, see Figure~\ref{fig:k64ising-gap}.

\subsection{Embedded $K_{64}$ MaxCut Instances}

In addition to the random Ising  models
we are also considering unweighted MaxCut problems on randomly generated subgraphs of $K_{64}$.
The corresponding Ising  objective for a graph $G = (V,E)$ is $\sum_{ij \in E} s_i s_j$ with $s \in \{-1, 1\}^V$, 
hence all Chimera edges get weight 1 and there is no $h$-field. 
The maximum cut size can then be obtained from an optimum solution $s^*$ by $-\frac{1}{2} (\sum_{ij \in E} s^*_i s^*_j - |E|)$.

For the transformation of the instances to Chimera Ising  instances we proceed as in the previous subsection. In order to obtain integer coefficients between $-10$ and $10$ we had to keep the density below $0.2$.
Therefore, we have chosen the edge probabilities $0.18,0.19,0.20$ for generating the random instances.
In total, 80 embedded instances were created out of 40 suitable graphs with 64 nodes each.
The embedding preserved the property of the non-existing $h$-field.

\begin{figure}[ht]
\begin{center}
\includegraphics[scale=.3]{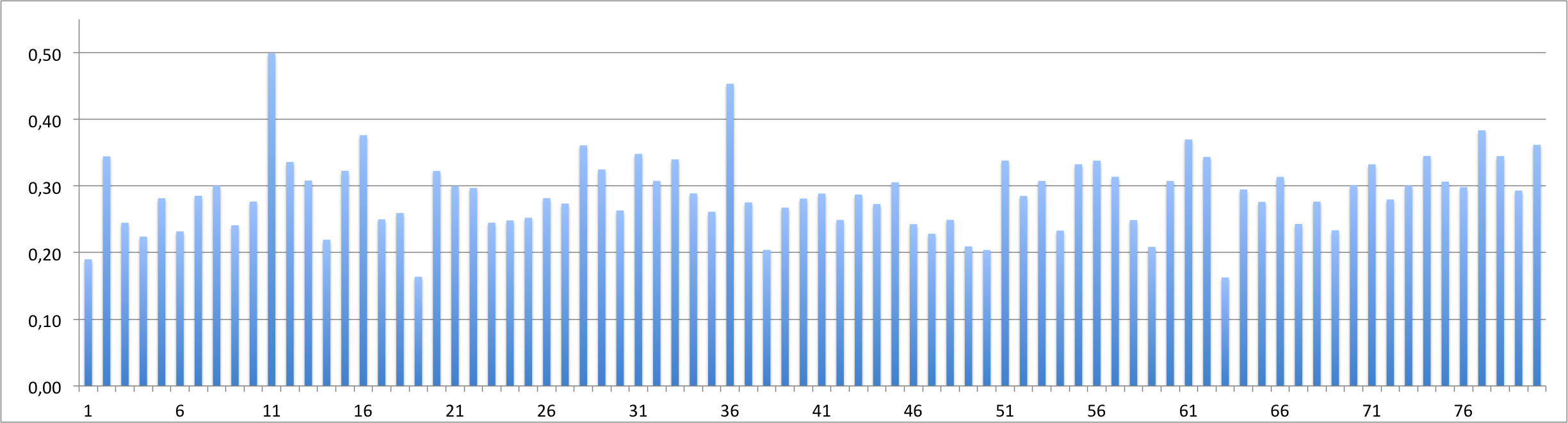} 
\caption{Relative deviations of the best \dwave\ energies from true ground state energies for the embedded $K_{64}$ MaxCut instances on $C_{16}$ in percent}
\label{fig:k64maxcut-gap}
\end{center}
\end{figure}

The only code that succeeded to find true ground states has been, quite surprisingly, \texttt{generalmaxcut}, but only when the time limits were given up entirely. The longest run took almost 2 cpu days while the shortest took only 1 cpu second. 
\dwave\ never finds a true ground state, and the energy differs from the ground state energy by at most $0.499$\% and by $0.288$\% on average%
, see Figure~\ref{fig:k64maxcut-gap}.

\subsection{Performance of the Selby  Exact Solver and the Selby  Heuristic}

We ran experiments with our implementation of the Selby  exact algorithm on $C_8$ instances. Indeed, the
times are comparable with those of the other exact
algorithms. Sometimes they are the best among them.

For all the instances for which the optimum solution is known, the
heuristic always found a ground state.  
For all McGeoch-Wang instances
but the 10th and the 13th, for all large Selby instances, and for large gate-assignment-like instances 3,
9, 10, 27, 28, and 29 we did not succeed in finding a mathematically
certified optimum solution, therefore we ran the heuristic with a
maximum time of 1 hour in order to generate a reliable ``best`` value.

For 11 instances of this set the Selby  heuristic did not find the best known
solution within the standard allowed time of 30~seconds, but when we
allowed 100~seconds instead, the heuristic found the best known
solution for all instances but one, namely the 10th large gate-assignment-like instance,
for which 6~minutes were necessary to match the best known solution.  

\subsection{Summary}

In the following tables, we provide some additional statistics and summarize our results. The rows correspond to the eleven instance sets

\begin{footnotesize}
\begin{center}
\begin{tabular}{ll}
\texttt{mgw-c8-439} & McGeoch-Wang instances on 439 qubits of $C_8$,\\
\texttt{mgw-c8-507} & McGeoch-Wang instances on 507 qubits of $C_8$,\\
\texttt{mgw-c16-2031} & McGeoch-Wang instances on 2031 qubits of $C_{16}$,\\
\texttt{rfr} & random instances with full range within granularity,\\
\texttt{selby-c8} & Selby  instances on $C_8$,\\
\texttt{selby-c16} & Selby  instances on $C_{16}$,\\
\texttt{mis} & maximum independent set instances,\\
\texttt{pga} & practical instances from gate assignment,\\
\texttt{lga} & large gate-assignment-like instances,\\
\texttt{k64ising} & embedded $K_{64}$ Ising  instances,\\
\texttt{k64maxcut} & embedded $K_{64}$ MaxCut instances.
\end{tabular}
\end{center}
\end{footnotesize}

The columns are

\begin{footnotesize}
\begin{center}
\begin{tabular}{lp{99mm}}
\textit{\#instances} & total number of instances,\\
\textit{\#nodes (min/max/average)} & number of nodes after transformation into MaxCut instances (there is an extra node for the field),\\
\textit{\#edges (min/max/average)} & number of edges after transformation into MaxCut instances,\\
\textit{\#opt-known} & number of instances for which a ground state is known due to one of the exact solvers,\\
\textit{\#Selby-best-100} & number of instances for which the Selby  heuristic found the best known solution within 100 seconds,\\
\textit{\#Selby-best-30} & number of instances for which the Selby  heuristic found the best known solution within 30 seconds,\\
\textit{\#DW-best} & number of instances for which \dwave\ found the best known solution,\\
\textit{\%DW-gap (max/average)} & relative gap between best known and \dwave\ energies in percent.\\
\end{tabular}
\end{center}
\end{footnotesize}

Table~\ref{table:probstats} gives some statistics on the eleven problem sets. 
Table~\ref{table:results} summarizes the performance of our exact algorithms as well as of the Selby  heuristic and \dwave. 

\begin{table}[!ht]
\begin{footnotesize}
\begin{center}
\begin{tabular}{|l|r|r|r|r|r|r|r|}
\cline{2-8}
\multicolumn{1}{r|}{} &
\textit{\#in-} & 
\multicolumn{3}{c|}{\textit{\#nodes}} & 
\multicolumn{3}{c|}{\textit{\#edges}}  \\
\cline{3-5}
\cline{6-8}
\multicolumn{1}{r|}{} &\textit{stances}&\textit{min}&\textit{max}&\textit{avg}&\textit{min}&\textit{max}&\textit{avg}\\
\hline
\texttt{mgw-c8-439}    &100   &    440    &   440&  440.00   &   1509     & 1535 &1521.70       \\
\texttt{mgw-c8-507}    &20    &   508     &  508 & 508.00     & 1951    &  1951 &1951.00         \\
\texttt{mgw-c16-2031}&20   &   2032   &   2032 &2032.00    &  7950    &  7950 &7950.00  \\
\texttt{rfr}                    &20   &   2032   &   2032 &2032.00   &   7541  &    7618 &7577.70                      \\
\texttt{selby-c8}          &20   &    507      & 507&  507.00     & 1314  &    1347 &1331.55               \\
\texttt{selby-c16}        &20    &  2031  &    2031& 2031.00  &    5399     & 5494& 5462.85          \\
\texttt{mis}                 &20   &   2032   &   2032 &2032.00  &    4411   &   4600 &4489.40                    \\
\texttt{pga}                 &1035     &    9     &   88  & 22.12    &    24     &  246  & 59.79                           \\
\texttt{lga}                  &30   &   502    &  1745 &1060.87  &    1636   &   5043 &3230.13                      \\
\texttt{k64ising}         &80    &  1184   &   1275& 1231.53    &  1852  &    2092 &1973.24            \\
\texttt{k64maxcut}     &80   &   1101   &   1226& 1175.03   &   1370  &    1577& 1482.28        \\
\hline
\end{tabular}
\caption{Statistics on the instances}
\label{table:probstats}
\end{center}
\end{footnotesize}
\end{table}

\begin{table}[!ht]
\begin{footnotesize}
\begin{center}
\begin{tabular}{|l|r|r|r|r|r|r|r|}
\cline{2-8}
\multicolumn{1}{r|}{} &
\textit{\#in-} & 
\textit{\#opt} & 
\textit{\#Selby} & 
\textit{\#Selby} & 
\textit{\#DW} & 
\multicolumn{2}{c|}{\textit{\%DW-gap}} \\
\cline{7-8}
\multicolumn{1}{r|}{} &\textit{stances}&\textit{-known}&\textit{-best-100}&\textit{-best-30}&\textit{-best}&\textit{max}&\textit{avg}\\
\hline
\texttt{mgw-c8-439}       & 100  & 100 &100  &100            & 95       &0.254    &0.013\\
\texttt{mgw-c8-507}       &  20  &  20  &20   &20              &20        &0.000    &0.000\\
\texttt{mgw-c16-2031}   &   20 &   2   &20   &20             &4          &0.201    &0.082\\
\texttt{rfr}                       &  20  &  20   &20  &20             & 0         &0.258    &0.137\\
\texttt{selby-c8}             &  20  &  20   &20    &20             &13        &0.165    &0.028\\
\texttt{selby-c16}           &   20 &   0    &20  &10              &0        &0.584    &0.396\\
\texttt{mis}                     &  20  &  20   &20  &20             & 4        &0.647    &0.257\\
\texttt{pga}                    &1035&1035 &1035  &1035          &1035   &0.000   &0.000\\
\texttt{lga}                     &  30   &  24   &29   &29              & 2        &0.347    &0.145\\
\texttt{k64ising}            &  80   &  80   &80   &80              &48        &0.134    &0.013\\
\texttt{k64maxcut}        &  80   &  80   &80  &80              &0         &0.499   &0.288\\
\hline
\end{tabular}
\caption{Performance of the Selby  heuristic and \dwave\ on the instancess}
\label{table:results}
\end{center}
\end{footnotesize}
\end{table}

The easiest instance sets for our exact approaches have been \texttt{pga}, \texttt{mis}, and \texttt{k64ising}, all of which could be solved within a few seconds. The reason for the  \texttt{pga} instances is their small size. On the other hand, the \texttt{mis} instances belong to the largest tested instances. Before transformation to Ising  instances on Chimera graphs, they are independent set instances on bipartite graphs that can be solved in polynomial time. Our approaches seem to be able to profit from this fact while the \dwave\ machine is not. 
For \dwave, these instances belong to the hardest tested instances.
This confirms observations made by Parekh  et al.~\cite{ParekhWS+16}. The \texttt{k64ising} instances have a nonzero field and belong to the sparsest instances in our study, which explains that \texttt{generalmaxcut} solves them very quickly.
Also for \dwave, these instances are not hard, although the ground state cannot be found for 40\% of the instances.

The instance sets of medium difficulty for our exact approaches have been \texttt{rfr}, \texttt{mgw-c8-439}, and \texttt{mgw-c8-507}. These were solved to optimality in a few minutes. Although the \texttt{rfr} instances belong to the largest tested instances, their weight distribution (weights of wider range) made \bc\ approaches easier. This is in line with observations that have been made in the context of solving Ising spin glass problems on grid graphs, where instances with a Gaussian distribution on the edge weights have turned out to be much easier than those with $\pm1$ edge weights. The McGeoch-Wang instance sets \texttt{mgw-c8-439} and \texttt{mgw-c8-507} belong to the smallest with a nonzero field.
They are also easy for \dwave.

The most difficult instance sets  for our exact approaches have been \texttt{selby-c8}, \texttt{k64maxcut}, \texttt{lga}, \texttt{mgw-c16-2031}, and \texttt{selby-c16}. The Selby  instances have been constructed to be very hard (zero field and difficult weight distribution) and they have kept their promise. Nevertheless, we have been able to solve all \texttt{selby-c8} instances to optimality due to their small size, but the \texttt{selby-c16} have been the hardest instances of the entire study, both for our exact algorithms as well as the \dwave\ machine that had the largest average gap for this set. 
With an edge/vertex ratio of only $1.26$ the \texttt{k64maxcut} instances are the sparsest instance set which might explain why \texttt{generalmaxcut} has been able to solve them all to optimality, yet other than the sparse \texttt{k64ising} instances they have a zero field, making them harder as already observed in the study of Dash and Puget~\cite{DashP15}. The \texttt{lga} instances have been constructed to be large versions of the \texttt{pga} instances, but they are much larger. Their edge/vertex ratio is roughly twice the one of the \texttt{k64} instances and their weight structure
makes them so hard that we have not been able to solve them all to provable optimality. The \texttt{mgw-c16-2031} instances are the largest and densest  of the entire study, and clearly beyond our current ability.

For  $1401$ of the $1445$ instances of this study, we have been able to compute proven ground states, in the remaining  $44$ cases our best solution has a proven high quality: For the 18 \texttt{mgw-c16-2031} instances for which the ground state energy remains unknown, the maximum relative energy deviations of our best solutions from the ground state energy are between $0.189$\% and  $0.894$\%. For the 20 \texttt{selby-c16} instances, the maximum relative deviations are between $1.266$\% and $3.982$\%, and for the 6 \texttt{lga} instances, the maximum relative deviations are between $0.015$\% and $0.092$\%.
 
\section{Conclusion}

When we started this study,
we conjectured ``What is hard for our mathematical optimization solvers in terms of running time is hard for the quantum annealer in terms of solution quality''. But the outcome decribed above shows that this is not always true. 

By quantum annealing  the \dwave\ computer realizes an extremely fast heuristic
for certain Ising  problems. It cannot be expected to deliver true optimum solutions, but the relative deviations will
usually be very small: The maximum relative deviation from the ground state energy is $0.647$\% over all $1445$ instances in this study. By spending a larger amount of computing time, solutions of this quality can also be
achieved by, e.g., classical simulated annealing, the Selby  heuristic (only applicable for Chimera graphs)
or general MaxCut heuristics. 

We observed that for $C_8$/$C_{16}$ pairs like \texttt{mgw-c8/mgw-c16} and \texttt{selby-c8/selby-c16} as well as for the pair \texttt{pga/lga}, the quality of the \dwave\ solutions deteriorates with increasing problem size.

The most surprising result is 
the remarkably excellent performance of the Selby  heuristic. Like \dwave, the Selby  heuristic cannot give an approximation quality guarantee, but the experiments show that it can produce very good solutions in short computation time, so it is an alternative to be taken seriously. 

If one needs to compute truely optimum solutions then general MaxCut solvers based on polyhedral combinatorics or semidefinite programming are the methods of choice. 

While our exaxt approaches have been tailored to the special architecture of Chimera graphs, we expect that similar approaches will also be possible for future architectures of \dwave\ machines.

\section{Acknowledgments}

The authors would like to thank NASA Quantum Artificial Intelligence Laboratory for many valuable discussions and the opportunity to use the \dwave\ 2000Q machine at NASA Ames. We gratefully acknowledge the funding from the European Union's Horizon 2020 research and innovation programme under the Marie Sk{\l}odowska-Curie grant agreement No 764759 \emph{MINOA Mixed-Integer Nonlinear Optimisation: Algorithms and Applications}.

\bibliographystyle{plain}
\bibliography{chimera-arxiv}

\end{document}